\documentclass[
    ,final            % use final for the camera ready runs
  ]
  {aipproc}

\layoutstyle{8x11double}

\newif\ifpreprint
\preprinttrue     % uncomment this line for preprint version

\begin{document}

\title{
\ifpreprint
\rightline{\normalsize\bf UASLP--IF--03--008}\vspace{0.2cm}
Review of Recent Results in Charm Physics\footnote{Invited talk the
9th International Conference on B-Physics
at Hadron Machines -- BEAUTY 2003,
Carnegie Mellon University, Pittsburgh, October 14-18, 2003.
Proceedings to be published by AIP.}
\else
Review of Recent Results in Charm Physics
\fi
}

\ifpreprint
\author{J\"urgen Engelfried}{
address={Instituto de F\'{\i}sica, Universidad Aut\'onoma de 
San Luis Potos\'{\i}, 
         San Luis Potos\'{\i} 78000, M\'exico
{\url{jurgen@ifisica.uaslp.mx}}
}}
\else
\author{J\"urgen Engelfried}{
address={Instituto de F\'{\i}sica, Universidad Aut\'onoma de 
San Luis Potos\'{\i}, 
         San Luis Potos\'{\i} 78000, M\'exico},
email={jurgen@ifisica.uaslp.mx}
}
\fi

\begin{abstract}
A biased review of recent results in charm physics is presented. New
results on $D^0-\overline{D^0}$ mixing, rare decays of $D^0$ and
$D^\pm$, scalar resonances in $D^+$ and $D_s$ decays, and new decay modes
and mass measurements in $\Lambda_c^+$, $\Xi_c^{+,0}$, $\Omega_c^0$, and
$\Xi_{cc}^+$ are discussed.
\end{abstract}

\maketitle

%%%%%%%%%%%%%%%%%%%%%%%%%%%%%%%%%%%%%%%%%%%%
%% MAINMATTER
%%%%%%%%%%%%%%%%%%%%%%%%%%%%%%%%%%%%%%%%%%%%

\section{Introduction}

In contrary to the last 5 years or so, were mostly ``traditional''
charm experiments
like E791, FOCUS, SELEX, WA89, WA92, 
CLEO, and H1/ZEUS published results about more ``traditional'' topics
like production, lifetimes, rare decays, and limits on 
$D^0-\overline{D^0}$ mixing, accompanied by a
small number of theory and phenomenology papers,
in the last year a shift in charm physics occurred.
New players like BaBar, Belle and CDF entered the field, new
charm states (doubly charmed baryons,
hidden double charm ($J/\Psi\,c\overline{c}$), $D_s^*$, $X(3872)$)
were discovered,
and the first pentaquark was observed.
All this triggered a 
large number of ``theory'' papers, pre- and post-dicting the 
spectroscopy and production of these new states.
In most of these papers a (back-)shift to the di-quark picture of
charmed hadrons can be observed.

We will present here a (biased)
selection of recent results in charm physics.  In several other talks 
at this conference charm results were shown.

\section{${\mathbf{D^0-\overline{D^0}}}$ Mixing}

\begin{table}
\begin{tabular}{lll}
\hline
Experiment&\multicolumn{1}{c}{Measurement}&\multicolumn{1}{c}{Reference}\\
\hline
Belle& $y_{CP}=(+1.15\pm 0.69\pm 0.38)\,\mbox{\%}$&\cite{Abe:2003ys}\\
%\hline
BaBar& $y_{CP}=(-0.8\pm 0.4\,^{+0.5}_{-0.4})\,\mbox{\%}$&
\cite{Aubert:2003pz}\footnote{also includes $D^0\to \pi^+\pi^-$}\\
%\hline
CLEO& $y_{CP} = (-1.2 \pm 2.5 \pm 1.4)\,
\mbox{\%}$&\cite{Csorna:2001ww}\footnote{also includes $D^0\to \pi^+\pi^-$}\\
%\hline
FOCUS& $y_{CP} = (3.42 \pm 1.39 \pm 0.74)\,\mbox{\%}$&\cite{Link:2000cu}\\
%\hline
E791& $y_{CP} = (0.8 \pm 2.9 \pm 1.0)\,\mbox{\%}$
&\cite{Aitala:1999dt}\footnote{Measured $\Delta\Gamma=(0.04\pm0.14\pm0.05)\,\mbox{ps}^{-1}$}\\
%\hline
BaBar& $-0.056 < y^\prime < 0.039 $ (95\% C.L.)&\cite{Aubert:2003ae}\\
%\hline
CLEO& $-0.058 < y^\prime < 0.01 $ (95\% C.L.)&\cite{Godang:1999yd} \\
\hline
\end{tabular}
\caption{Recent measurements of $CP$ violation observables in the
$D^0$ system.}
\label{cp}
\end{table}
The usual observable for ${CP}$ violation in the charm system
is the lifetime difference 
between $D^0\to K^-K^+$
and $D^0\to K^-\pi^+$, defined as
$y_{CP} = {{\tau(K^-\pi^+)}/{\tau(K^-K^+)}} - 1$, predicted in 
the Standard Model to $y_{CP}\sim10^{-3}$.
Another possible analysis is the ``wrong-sign''
Double Cabbibo Suppressed $D^0\to K^+\pi^-$, with the
observable $y^\prime$.
Recent results where published by Belle~\cite{Abe:2003ys} and
BaBar~\cite{Aubert:2003pz}, and are compared
with previous results in table~\ref{cp}.

All measurements are compatible with $0$, e.g.\ no ${CP}$ violation
was observed yet in the charm system.

\section{Rare Decays of {$\mathbf{D^0}$} and {$\mathbf{D^\pm}$} Mesons}
FOCUS observed the rare decay $D^0\to K^-K^-K^+\pi^+$ with 
a yield of $132\pm19$ events, and measured the relative
branching ratio to
${{\Gamma(D^0\to K^-K^-K^+\pi^+)}/{\Gamma(D^0\to K^-\pi^-\pi^+\pi^+)}}= 
0.00257\pm0.00034\pm0.00024$~\cite{Link:2003pt}. Resonant 
Resonant substructures with $\Phi$ and $\overline{K^*(892)^0}$ are
dominant.

Belle observed  $D^0\to \phi\pi^0$, $\phi\eta$, and 
$\phi\gamma$~\cite{Abe:2003yv}.

CLEO performed a Dalitz plot analysis of
$D^0\to \pi^-\pi^+\pi^0$,
and studied $D^0\to K_s\eta\pi^0$~\cite{Dubrovin:2003jt}. CLEO also
observed the Cabbibo suppressed decays
$D^+\to\pi^+\pi^0$, $K^+\overline{K^0}$, and $K^+\pi^0$~\cite{Arms:2003ra},
and the measured Branching Ratios are shown in table~\ref{dplus}.
\begin{table}
\begin{tabular}{ll}
\hline
${\cal B}(D^+\to\pi^+\pi^0)$&$(1.31\pm0.17\pm0.09\pm0.09)\cdot10^{-3}$\\
${\cal B}(D^+\to K^+ \overline{K^0})$&
$(5.24\pm0.43\pm0.20\pm0.34)\cdot10^{-3}$\\
${\cal B}(D^+\to K^+ \pi^0)$&$ < 4.2\cdot10^{-4} \mbox{~~(90\% C.L.)}$\\
\hline
\end{tabular}
\caption{Branching Ratios for $D^+$ decays, measured by
CLEO~\cite{Arms:2003ra}.}
\label{dplus}
\end{table}

FOCUS studied di-muon decays for $D^+$ and $D_s^+$~\cite{Link:2003qp},
and obtained new limits on these modes.

A new player in the field, CDF, set a limit
for $D^0\to\mu^+\mu^-$ at $<2.5\cdot10^{-6}$~\cite{Acosta:2003ag}.

\section{Scalar Resonances in {$\mathbf{D^+}$} and 
{$\mathbf{D_{\lowercase{s}}^+}$} Decays}

Since a few years E791 is studying the modes 
$D^+\to K^-\pi^+\pi^+$,
$D^+\to \pi^-\pi^+\pi^+$, and
$D_s^+\to \pi^-\pi^+\pi^+$. 
To explain the resonant substructures in the decays, they 
need to include two scalar resonance, one for 
$K\pi$ (the $\kappa$) with mass $(797\pm19\pm43)\,\mbox{MeV}/c^2$ and 
width $(410\pm43\pm87)\,\mbox{MeV}/c^2$, and a second in 
$\pi\pi$ (the $\sigma$) with mass $(478_{-23}^{+24}\pm17)\,\mbox{MeV}/c^2$ and
 width $(324_{-40}^{+42}\pm21)\,
\mbox{MeV}/c^2$~\cite{Aitala:2002kr,Aitala:2000xu,Bediaga:2003wv}.

\section{The {$\mathbf{D_{\lowercase{s}}}$} System}
On April 12, 2003,  BaBar announced the observation of a narrow resonance,
decaying to $D_s\pi^0$, at $2.32\,\mbox{GeV}/c^2$~\cite{Aubert:2003fg}.
Shortly after, CLEO not only confirmed the observation, but observed
an additional resonance,
decaying to $D_s^*\pi^0$~\cite{Besson:2003cp,Stone:2003cu}.
During the summer conferences, Belle confirmed both
observations~\cite{Krokovny:2003zq,Abe:2003jk}.
The most likely nature of these states are excited $D_s$ mesons; the search
for similar states in the $D^0$ and $D^\pm$ system already started.
More details can be found in these proceedings~\cite{Wang}.

%\section{Charmed Baryons: The {$\mathbf{\Lambda_c^+}$} and
\section{Charmed Baryons: The {${\Lambda_{\lowercase{c}}^+}$} and
%{$\mathbf{\Sigma_{\lowercase{c}}^{0,++}}$}}
{${\Sigma_{\lowercase{c}}^{0,++}}$}}
CLEO reports the observation of
the $\Lambda_c^+\to \Lambda\pi^+\pi^+\pi^-\pi^0$
decay~\cite{Cronin-Hennessy:2002we}, 
with ${\cal B}=(1.79\pm0.47\pm0.43)\,\mbox{\%}$, while
most of the decays 
happen via $\Lambda_c^+\to\Lambda\omega\pi^+$.

CLEO also measured the masses and widths of $\Sigma_c^{++}$ and 
$\Sigma_c^0$~\cite{Artuso:2001us}
(the results are shown in table~\ref{sigmac}),
\begin{table}
\begin{tabular}{ll}
\hline
$M(\Sigma_c^{++})-M(\Lambda_c^+)$&$(167.4\pm0.1\pm0.2)\,\mbox{MeV}/c^2$\\
$M(\Sigma_c^{0})-M(\Lambda_c^+)$&$(167.2\pm0.1\pm0.2)\,\mbox{MeV}/c^2$\\
$\Gamma(\Sigma_c^{++})$&$(2.3\pm0.2\pm0.3)\,\mbox{MeV}/c^2$\\
$\Gamma(\Sigma_c^{0})$&$(2.5\pm0.2\pm0.3)\,\mbox{MeV}/c^2$\\
\hline
\end{tabular}
\caption{Masses and Width for $\Sigma_c^{++}$ and $\Sigma_c^0$ as
measured by CLEO~\cite{Artuso:2001us}.}
\label{sigmac}
\end{table}
updating previous results on the masses from E791~\cite{Aitala:1996cy}.

%\section{Charmed Baryons: The {$\mathbf{\Xi_c^+}$} and {$\mathbf{\Xi_c^0}$}}
\section{Charmed Baryons: The {${\Xi_{\lowercase{c}}^+}$} and 
{${\Xi_{\lowercase{c}}^0}$}}
FOCUS measured several new decay modes of the $\Xi_c^+$
and re-measured some previously
observed ones. A summary is given in table~\ref{xic}.
\begin{table}
\begin{tabular}{cccc}
\hline
Decay Mode&FOCUS~\cite{Link:2003cd}&CLEO\cite{Bergfeld:1996qn}
    &SELEX~\cite{Jun:1999gn} \\
\hline
${{\Gamma(\Xi_c^+\to\Sigma^+ K^-\pi^+)}\over
{\Gamma(\Xi_c^+\to\Xi^-\pi^+\pi^+)}}$&
$0.91\pm0.11\pm0.04$&$1.18\pm0.26\pm0.17$&$0.92\pm0.20\pm0.07$\\ \\
${{\Gamma(\Xi_c^+\to\Sigma^+ K^+K^-)}\over
{\Gamma(\Xi_c^+\to\Sigma^+ K^-\pi^+)}}$&
$0.16\pm0.06\pm0.01$& & \\ \\
${{\Gamma(\Xi_c^+\to\Lambda^0 K^-\pi^+\pi^+)}\over
{\Gamma(\Xi_c^+\to\Xi^-\pi^+\pi+)}}$&
$0.28\pm0.06\pm0.06$&$0.58\pm0.16\pm0.07$& \\ \\
${{\Gamma(\Xi_c^+\to\Omega^- K^+\pi^+)}\over
{\Gamma(\Xi_c^+\to\Xi^-\pi^+\pi+)}}$&
$0.07\pm0.03\pm0.03$& & \\ \\
${{\Gamma(\Xi_c^+\to\Sigma^*(1385)^+\overline{K^0})}\over
{\Gamma(\Xi_c^+\to\Xi^-\pi^+\pi+)}}$&
$1.00\pm0.49\pm0.24$& & \\ \\
\hline
\end{tabular}
\caption{Relative Branching Ratios for $\Xi_c^+$.}
\label{xic}
\end{table}
FOCUS also includs  upper limits for resonances in these decay modes.

CLEO obtained a new measurement of the $\Xi_c^+$ lifetime, 
$\tau(\Xi_c^+)=(503\pm47\pm18)\,\mbox{fs}$~\cite{Mahmood:2001em}.

CLEO also reports the first observation of the 
$\Xi_c^0\to p K^-K^-\pi^+$ decay~\cite{Danko:2003va}, with a relative
branching ratio of 
${{{\cal B}(\Xi_c^0\to p K^-K^-\pi^+)}/
{{\cal B}(\Xi_c^0\to\Xi^-\pi^+)}} = 0.35\pm0.08\pm0.05$.
In this decay they see evidence for a
resonant $\overline{K^*(892)^0}$ substructure.

%\section{Charmed Baryons: The {$\mathbf{\Omega_c^0}$}}
\section{Charmed Baryons: The {${\Omega_{\lowercase{c}}^0}$}}
Evidence for the  $\Omega_c^0$ in $e^+e^-$ interactions was reported 
long time ago by ARGUS~\cite{Albrecht:1992xa}, and now first
CLEO~\cite{Cronin-Hennessy:2000bz}
and recently Belle~\cite{Ammar:2002pf,BelleOmega:2003} confirm
this observation. Both measure
the mass of the $\Omega_c^0$
(Belle: $(2693.9\pm 1.1 \pm 1.4)\,\mbox{MeV}/c^2$,
CLEO: $(2694.6\pm 2.6\pm 1.9)\,\mbox{MeV}/c^2$) significantly different
from the PDG2000: $(2704\pm 4)\,\mbox{MeV}/c^2$~\cite{Groom:2000in}.
Both observe the mode $\Omega_c^0\to \Omega^-\pi^+$ and
$\Omega_c^0\to \Omega^-e^+\nu$, and Belle observes in
addition the semileptonic muon mode.

%\section{Doubly Charmed Baryons: The {$\mathbf{\Xi_{cc}^+}$}}
\section{Doubly Charmed Baryons: The {${\Xi_{\lowercase{cc}}^+}$}}
The SELEX experiment reported the first observation of
a member of the doubly charmed baryon family, the $\Xi_{cc}^+$, in the
decay mode $\Xi_{cc}^+\to\Lambda_c^+ K^- \pi^+$~\cite{Mattson:2002vu}.
Further work on different decay modes is ongoing.

%%%%%%%%%%%%%%%%%%%%%%%%%%%%%%%%%%%%%%%%%%%%%%%%
%% BACKMATTER
%%%%%%%%%%%%%%%%%%%%%%%%%%%%%%%%%%%%%%%%%%%%%%%%

\begin{theacknowledgments}
I would like to thank John Cumalat and Erik Gottschalk from FOCUS, 
Sheldon Stone and  JC Wang from CLEO,
Jeff Appel from E791, 
Masa Yamauchi and Karim Trabelsi from Belle,
and Livio Lanceri from BaBar for providing me with information and figures
from their respective experiments.

I would also like to thank the organizers of the conference for the
opportunity to give this presentation.
\end{theacknowledgments}

%% If the bibliography is produced without BibTeX comment out the

\bibliographystyle{aipproc}   % if natbib is available

\bibliography{Beauty2003_jurgen}

\end{document}